# The Period-Modulated Harmonic Locked Loop (PM-HLL): A low-effort algorithm for rapid time-domain multi-periodicity estimation

Volker Hohmann[1,2,3]

[1] Department of Medical Physics and Acoustics, University of Oldenburg, Oldenburg, Germany

[2] Hörzentrum Oldenburg gGmbH, Oldenburg, Germany

[3] Cluster of Excellence Hearing4all Oldenburg, Germany

**Corresponding author:**

Prof. Dr. Volker Hohmann

Department of Medical Physics and Acoustics and Cluster of Excellence Hearing4all

University of Oldenburg

D-26111 Oldenburg

Germany

Email: volker.hohmann@uol.de

Phone: +49 441 798 5468

**Abstract:**

Many speech and music analysis and processing schemes rely on an estimate of the fundamental frequency $f_0$ of periodic signal components. Most established schemes apply rather unspecific signal models such as sinusoidal models to the estimation problem, which may limit time resolution and estimation accuracy. This study proposes a novel time-domain locked-loop algorithm with low computational effort and low memory footprint for $f_0$ estimation. The loop control signal is directly derived from the input time signal, using a harmonic signal model. Theoretically, this allows for a noise-robust and rapid $f_0$ estimation for periodic signals of arbitrary waveform, and without the requirement of a prior frequency analysis. Several simulations with short signals employing different types of periodicity and with added wide-band noise were performed to demonstrate and evaluate the basic properties of the proposed algorithm. Depending on the Signal-to-Noise Ratio (SNR), the estimator was found to converge within 3-4 signal repetitions, even at SNR close to or below 0dB. Furthermore, it was found to follow fundamental frequency sweeps with a delay of less than one period and to track all tones of a three-tone musical chord signal simultaneously. Quasi-periodic sounds with shifted harmonics as well as signals with stochastic periodicity were robustly tracked. Mean and standard deviation of the estimation error, i.e., the difference between true and estimated $f_0$, were at or below 1 Hz in most cases. The results suggest that the proposed algorithm may be applicable to low-delay speech and music analysis and processing.

# 1. Introduction

Oscillations are common processes in natural and technical systems, leading to (quasi-) periodicity constituting an important information-bearing feature of the signals emerging from such systems. Examples of natural signals are human speech, music, animal songs and echolocation cries. Technical examples are oscillating mechanical or electrodynamic systems. Detecting and processing (quasi-) periodicity is therefore a key element of psychoacoustic and physiologic auditory systems modeling (e.g., [1][2]), speech and music analysis and processing (e.g., [3][4]), radio communication systems (e.g., [5]) as well as of power electronics systems (e.g., [6]). This study derives a locked-loop algorithm for fast detection and estimation of (quasi-) periodicity emerging in time signals. It is motivated by auditory modeling approaches and adopts several processing principles commonly used in audio signal processing. A basic evaluation of the algorithm is provided using tests signals typical for audio applications, demonstrating its potential for widespread application in audio signal processing.

Signal periodicity emerges in many physical systems under the influence of internal delay, such as delayed feedback or coupling [7]. Therefore, modeling emerging periodicity often includes time-delayed coupling as well, examples being correlation analysis, differential equations and finite difference equations (i.e., filtering). To detect and estimate periodicity in the time domain, Phase-Locked Loops (PLL) are often used [8]. However, most PLLs do not include internal time delays corresponding to the periodicity to be detected, and are therefore best adapted to detecting sinusoidal signals by design. An exception is the Delay-Locked-Loop (DLL) [9][10], which uses internal delays to detect and estimate periodic signals of rectangular shape optimally. DLLs are used for binary data and clock signal regeneration [11] as well as for synchronizing power grids [12]. However, they are not suited for (quasi-) periodic signals of an arbitrary waveform shape. Several locked-loop techniques were proposed to estimate such generic periodic signals in audio signal processing [13][14], but these techniques use several PLLs to estimate the fundamental and the harmonics of the periodic signal in parallel, which makes them computationally rather expensive and implies the underlying component-wise signal model to be sinusoidal rather than periodic. The contribution of this study is the combination of the locked-loop principle with delay-based superposition and cancellation mechanisms known from auditory systems modeling [15][16][17] to derive a novel delay-based locked loop, the Period-Modulated Harmonic Locked Loop (PM-HLL). The PM-HLL is explicitly tailored to modeling (quasi-) periodic signals of arbitrary waveform shape and can estimate emerging periodicity quickly with high accuracy and low memory footprint. To the author's knowledge no other locked-loop algorithm has been proposed that is directly controlled by a periodic signal component emerging in the input signal.

The remainder of the paper is organized as follows: First, the PM-HLL is introduced in detail, including a thorough discussion of its properties, leading to a set of testable hypotheses as a basis for simulation design. Then, simulation results are presented and discussed. Although the algorithm is motivated by auditory models and was evaluated with signals in the audio frequency range, results generalize towards frequency-transposed versions of these signals, because the PM-HLL is scale-invariant by design. Finally, potential applications of the PM-HLL in several fields are discussed. Matlab scripts generating the simulations and figures of this study are available at [18].

## 2. Methods

*Algorithm outline*

The PM-HLL is a time-domain locked-loop algorithm that adapts and locks to a periodic signal component with time-varying fundamental frequency $f_0(t)$ emerging in its input signal $x(t)$. Once locked to the periodic component, the PM-HLL adapts its internal **PM-HLL oscillator frequency** $f_c(t)$ over time to the fundamental frequency of the periodic component, i.e., the time course of the PM-HLL oscillator frequency is an estimate of the fundamental frequency: $f_c(t) = \hat{f}_0(t)$. $T_c(t) = 1/f_c(t)$ is an estimate of the signal's fundamental period $T_0(t)$, and is called the **PM-HLL oscillator period**.

In addition to the fundamental frequency, the PM-HLL estimates the **Harmonic-to-Noise Ratio** $HNR(t)$, i.e., the time-varying ratio between the power of the periodic component at $f_0(t)$ and the power of all other components present in the signal at time $t$. Locking of the PM-HLL to a periodic component is indicated by the HNR being above 0 dB, which is the expected value of the HNR for pure noise signals. The higher the HNR, the higher is the saliency of the periodic component the PM-HLL is locked to.

The PM-HLL only adapts and locks to a periodic signal component emerging in the input signal, if its fundamental frequency lies in a well-defined **PM-HLL catch range** around the current value of the PM-HLL oscillator frequency. In the absence of a periodic component in the catch range, the PM-HLL oscillator frequency $f_c(t)$ meanders freely with the HNR estimate being low, indicating that the PM-HLL is not locked. Note that the catch range follows the PM-HLL-oscillator frequency. If the fundamental frequency of the component the PM-HLL is locked to jumps out of the current catch range, the PM-HLL loses track.

In an application, several instances of the PM-HLL can be run in parallel or serially, each starting at a different initial PM-HLL oscillator frequency $f_{c,0}$, to scan a signal for the presence of periodic components in a certain pre-defined range of fundamental frequencies. For example, to scan a full octave, it can be divided into 12 semitones, and one PM-HLL instance is started in each semitone range with its PM-HLL oscillator frequency confined to that range.

In the following, the signal model and theory behind the PM-HLL algorithm are discussed, followed by the specifics of how it is implemented in this study.

*Signal Model and PM-HLL Algorithm*

The signal model underlying the PM-HLL is a periodic signal with time-varying fundamental frequency $f_0(t)$, computed as the derivative of the instantaneous phase $\varphi(t)$, and a finite number $N_p$ of harmonics:

$$x(t) = \sum_{n=1}^{N_p} a_n \sin\left(n\varphi(t) - \varphi_{0,n}\right)$$
$$f_0(t) = \frac{1}{2\pi} \frac{d\varphi(t)}{dt}$$
(1)

The amplitude $a_n$ and phase $\varphi_{0,n}$ of the single harmonics are not considered, i.e., only the signal periodicity emerging at the fundamental period $T_0 = 1/f_0$ drives adaptation and locking. Because uncontrolled amplitudes and relative phases of the harmonic components lead to an uncontrolled shape of the signal waveform, the PM-HLL cannot use a reference signal as a modulator for deriving a loop control signal, as it is the case for standard locked-loop algorithms [8]. Instead, the signal model (1) is explicitly used to define a period-synchronous modulation filter, which, when applied to the input signal, yields the phase-based control signal required to adapt the loop[1]. Although the PM-HLL has no intrinsic reference oscillator, the fundamental frequency estimate $f_c(t) = \hat{f}_0(t)$ is called the PM-HLL oscillator frequency throughout this work, as introduced above.

The remainder of this section describes the details of the PM-HLL algorithm, in particular period-stabilization and period-synchronous modulation filtering. Algorithm properties are discussed in some depth to be able to formulate testable hypotheses as a basis for simulation design.

*** Place Fig. 1 here ***

To stabilize the PM-HLL control loop and to compute *period-synchronous signal features*[2], the modeling concept of *strobed temporal integration* [17] is adopted. So-called *strobe points* are generated once per signal period in this concept. Whereas strobe points are computed from the input signal in [17], specifically from the signal envelopes in frequency subbands, the PM-HLL generates strobe points from its oscillator frequency $f_c$, which is loop-controlled by the input signal. Figure 1 shows the principle of strobe-point generation, assuming that $f_c$ increases linearly from 50Hz to 150Hz in the time interval 0ms to 100ms (plotted in blue, left scale). The corresponding phase function

---

[1] The period-synchronous modulation filter is specific to the PM-HLL algorithm, which is why its acronym includes "Period-Modulated" (PM). The term HLL was coined before for locked-loops that track harmonic signals [14], which is also the case for the PM-HLL.
[2] The term *pitch-synchronous* is often used in speech research, see, e.g., [19]. Because pitch is a perceptual quantity, however, this term is avoided to make clear quantities estimated here refer to physical quantities.

$$\Phi(t) = \int_0^t f_c(t')dt', \qquad (2)$$

normalized by $2\pi$ and taken modulo 1, is plotted in red (right scale). Whenever the phase wraps, a strobe point is set, indicated by black bars in Figure 1. The procedure yields a monotonically increasing sequence of strobe-point temporal positions

$$T_s(t) = \{t_{s,1}, t_{s,2}, t_{s,3}, \ldots, t_{s,n}\}, \qquad (3)$$

where $t_{s,1}$ denotes the temporal position of the first strobe point found, and $t_{s,n}$ denotes the position of the last strobe point elicited before the current time $t$. The length of the sequence increases with time. The time span between subsequent strobe points indicates the stabilized period. Averaging signal features such as fundamental frequency, signal-power or HNR across the time span between strobe points yields the corresponding period-synchronous (or: period-stabilized) features. Furthermore, strobed integration is applied by computing a *stabilized image* (called stabilized auditory image (SAI) in [17]) of a signal $s(t)$ by

$$\begin{aligned}
t' &= t - t_{s,n} \\
b_{SI}(t') &= lp_\tau\left(s(t_{s,n}+t'), s(t_{s,n-1}+t'), s(t_{s,n-2}+t'), \ldots, s(t_{s,n-N}+t')\right), \\
y_{SI}(t) &= b_{SI}(t')
\end{aligned} \qquad (4)$$

where $t$ denotes the current time instant, $t'$ denotes the time elapsed since the last strobe point, and $lp_\tau(\cdot)$ denotes lowpass filtering with $\tau$ being the time constant[3]. $s(t)$ may be the input signal $x(t)$ itself, or a filtered version of it (see below). Equation (4) means that the signal segment starting at each strobe point is continuously averaged into a signal buffer $b_{SI}$, which enhances the waveform of a periodic signal that emerges at the current stabilized period: Signal components that match the current stabilized period superpose constructively, whereas all other components average out ("noise"). According to (4), each entry in the signal buffer is updated once per period, and the buffer entry $b_{SI}(t')$ updated at time $t$ is copied to an ongoing period-stabilized output signal $y_{SI}(t)$[4].

---

[3] Note that the filter order $N$ of the lowpass filter determines the number of sampling points of the signal needed for the filtering. If a 1st-order IIR filter is used, only the current signal sample $s(t_{s,n}+t') \equiv s(t)$ is used.

*** Place Fig. 2 here ***

The main algorithmic novelty in this study is the proposed *period-synchronous modulation filter*, which loop-controls the adaptation of the PM-HLL oscillator frequency $f_c$. Figure 2 shows the transfer functions in the range between $f_c/2$ and $3f_c/2$ of the filters used in the proposed modulation filter (amplitude: blue curves, left scale; normalized phase: red curves, right scale). Two period synchronous filters are computed in parallel, a period-constructive (solid lines) and a period-suppressive (dashed lines) comb filter:

$$\begin{aligned} y_p(t) &= x(t) + x(t-T_c) \\ y_m(t) &= x(t) - x(t-T_c) \end{aligned} \quad (5)$$

where $T_c = 1/f_c$ is the current PM-HLL oscillator period and $y_p$, $y_m$ are the outputs of the period-constructive and –suppressive comb filters. If the fundamental frequency $f_0$ of the signal matches the PM-HLL oscillator frequency $f_c$, the output of the period-constructive filter is twice the input, whereas the output of the period-suppressive filter vanishes. If $f_0$ is lower (indicated as $f_{0,1}$) or higher (indicated as $f_{0,2}$) than $f_c$, the ratio between the filter output amplitudes $|y_p(t)/y_m(t)|$ decreases, and the phase difference between $y_p(t)$ and $y_m(t)$ is +90° or −90°, respectively. The PM-HLL uses this phase difference as its loop control signal: If the phase difference is positive, the PM-HLL oscillator frequency is increased by some factor, and it is decreased by some factor if the phase difference is negative. The PM-HLL control loop sensitivity is expected to be higher than for DLL designs due to the abrupt phase flip at the locking point. Several options exist for computing the time-varying phase difference, one of which is using the sign of the instantaneous frequency of the complex-valued signal $y_p(t)+iy_m(t)$. The filter outputs $y_p(t)$ and $y_m(t)$ may be period-stabilized according to Equation (4) prior to computing the phase difference, and/or the phase difference may be period-stabilized by taking a time-average across a small fraction of the current PM-HLL oscillator period $T_c$ before taking the sign. For the specific implementation of the PM-HLL used in this study see the next sub-section.

---

[4] Note also that a period-synchronous spectral profile of the signal can be computed from one period (between two subsequent strobe points) of the ongoing period-stabilized output signal $y_{SI}(t)$ by, e.g., linear prediction.

The HNR at the current PM-HLL oscillator frequency is estimated by

$$HNR(t) = \frac{\left\langle |y_p(t)|^2 \right\rangle_\tau}{\left\langle |y_m(t)|^2 \right\rangle_\tau}, \tag{6}$$

where $t$ denotes the current time instant and $\langle \cdot \rangle_\tau$ denotes a running temporal average with a time constant of $\tau$. The HNR may be period-stabilized by using the stabilized images of the comb filter outputs in the nominator and denominator and/or by adapting the time constant to be a multiple of the current PM-HLL oscillator period $T_c$.

Figure 2 shows a single period of the periodic comb filter transfer functions, centered at the PM-HLL oscillator frequency $f_c$. If the signal contains only the fundamental frequency component, the PM-HLL catch range is $\left[f_c(1-1/2), f_c(1+1/2)\right]$. Beyond this range, the phase difference flips, and the control signal becomes ambiguous. If the signal contains higher harmonics up to harmonic number $N_p$, the frequency difference $f_c - f_0$ at the fundamental increases to $N_p \cdot (f_c - f_0)$ at the highest harmonic. This means that the PM-HLL catch range reduces to

$$\Delta f_c = \left[f_c \cdot \left(1 - \frac{1}{2N_p}\right), f_c \cdot \left(1 + \frac{1}{2N_p}\right)\right] \approx \left[\frac{f_c}{\left(1 + \frac{1}{2N_p}\right)}, f_c \cdot \left(1 + \frac{1}{2N_p}\right)\right]. \tag{7}$$

As an example, the catch range for $N_p = 7$, which may be a typical choice for lowpass-filtered speech, is $\left[f_c\left(1 - \frac{1}{14}\right), f_c\left(1 + \frac{1}{14}\right)\right] \approx \left[\frac{f_c}{1.07}, 1.07 f_c\right]$, i.e., it corresponds to a semitone in each direction (factor $2^{1/12} \approx 1.06$). Thus, a bank of about six PM-HLL instances is required to monitor a full octave. If the highest harmonic is $N_p = 11$, which may be a typical choice for auditory models, the catch range is reduced, and about ten PM-HLL instances are required per octave. Note also that all PM-HLL instances can share a single delay line for computing the comb filters (5).

Note that a PM-HLL may still converge to periodic signals outside its catch range, because the higher harmonics that go beyond the non-ambiguity range may not fully destroy the phase difference signal, depending on the relative amplitude of the lower and the higher harmonics. To ensure convergence in the case of an emerging periodic component in the catch range, a lowpass shelving filter at $N_p \cdot f_c$ may be used to suppress ambiguous harmonics potentially present in the input. The implementation used in this study does not include such a filter.

Note that random noise components added to the periodic component contribute at random to the positive and negative phase regime of the period-synchronous modulation filter. The phase-based control signal thus may have a larger variation across time, but the noise does not bias its mean value. It is therefore expected that the PM-HLL is rather insensitive to additive noise. Furthermore, a somewhat counter-intuitive expectation is that increasing the noise bandwidth may improve the control-signal statistics, because the balance of components falling in the positive and in the negative phase regime may be improved. On the other hand, a control system that uses the comb-filter amplitude ratio for adaptation by, e.g., maximizing the HNR according to Equation (6), would need a rather long time constant to estimate the HNR gradient, at least in the presence of noise.

Note that small deviations from the harmonic signal model (1), e.g., by mistuning of some or all harmonics or by stochastic modulation, lead to quasi-periodic signals, which may not fully destroy the phase-based control signal. It is therefore expected that the PM-HLL also converges in the case of small deviations from the signal model.

Note also that all frequency components below $f_c/2$ lie in the positive phase regime, erroneously driving the PM-HLL towards a higher oscillator frequency. Depending on the strength of these components, a highpass shelving filter at $f_c/2$ may be needed to suppress these components. The implementation used in this study does not include such a filter.

*PM-HLL Implementation*

The digital implementation of the PM-HLL [18] used in this study closely follows the algorithm description given above. It includes the following processing steps:

1. Sample the input signal $x(t)$ at a sampling frequency $f_s = 5$ kHz.

2. Initialize an instance of the PM-HLL with an initial PM-HLL oscillator frequency $f_{c,0} \equiv f_c(t=0)$, which was chosen differently for the different simulations (cf. next section). $f_c$ was set to a minimum of 96Hz (80Hz in simulation VII), adaptation was limited to this lower value. A maximum number of harmonics $N_p = 7$ was set to define the catch range[5].

3. Compute the two comb-filter outputs $y_p$ and $y_m$ according to Equation (5) from the input signal. If the current PM-HLL oscillator period $T_c$ lies between two sampling points of the signal delay line, linear interpolation of sample values is applied.

4. Compute the stabilized image $y_{p,SI}$ and $y_{m,SI}$ of the two comb-filter outputs according to Equation (4).

5. Compute the HNR according to Equation (6) from the stabilized image $y_{p,SI}$ and $y_{m,SI}$ of the two comb-filter outputs.

6. Compute the control signal $cs(n)$ by computing the phase difference between subsequent samples of the complex-valued signal $y_{p,SI} + i y_{m,SI}$ and averaging the phase difference across time:

---

[5] Note that the catch range is not an explicit processing parameter of the algorithm. It defines the expected range of frequencies of a periodic component, which the PM-HLL instance will adapt and lock to.

$$\begin{aligned} c(n) &= y_{p,SI}(n) + i y_{m,SI}(n) \\ adc(n) &= \arg\left(c(n) \cdot c^*(n-1)\right) \\ cs(n) &= \langle adc(n) \rangle_\tau \end{aligned} \qquad (8)$$

7. Adapt the PM-HLL oscillator frequency $f_c$ at each sample by multiplication with (control signal negative) or division by (control signal positive) a factor. The factor is chosen such that the PM-HLL adapts to an emerging periodic component in its expected catch range (here: about a semitone) within two signal repetitions, i.e., within three times the PM-HLL oscillator period $T_c$ from the onset of the periodic component.

All running temporal averaging filters were implemented as 1st-order IIR lowpass filters with time constants defined as a multiple of the current PM-HLL oscillator period $T_c$. This means that all parameters controlling the PM-HLL relate to $T_c$, rendering the PM-HLL fully scalable with $T_c$. Time constants were $1.0 \cdot T_c$ for computing the stabilized image of the comb-filter outputs (processing step 4), $0.5 \cdot T_c$ for the computation of the HNR (processing step 5) and $0.1 \cdot T_c$ for the phase difference averaging (processing step 6). The HNR was further smoothed with a time constant of $0.05 \cdot T_c$. Thus, time constants were in the order of the magnitude of the oscillator period $T_c$ or below, ensuring quick adaptation of the loop. Informal tests showed that the PM-HLL convergence is robust against some variation in the time constants, but confirmed the expectation that the control loop starts oscillating if the time constants are increased to a small multiple of the oscillator period. Time constants were updated at each sample, but could also be updated once per period, i.e., when a strobe point is elicited.

The PM-HLL processing is performed sample by sample, i.e., it is a low-delay online algorithm. The adaptation time is expected to be about three times the period of an emerging periodic component and a fraction of this period once the loop is locked to a periodic component with continuously varying fundamental frequency. The memory footprint is mainly determined by the maximum oscillator period, according to Equation (5). At a minimum fundamental frequency to be tracked of 50 Hz and a sampling frequency of 5 kHz, e.g., the delay line is 100 samples long. Furthermore, a small number of filter states and input and output samples need to be stored. An autocorrelation function testing the same range of delays needs a delay line of the same length, but, being the inverse Discrete Fourier Transform of the power spectrum, uses implicitly a less specific sinusoidal signal model. The PM-HLL requires a couple of operations per sample, as described above, and thus is low in computational effort. A prior frequency analysis (short-time Fourier transform or filterbank) is not required.

## *Simulations*

By design and parameterization of the PM-HLL, several hypotheses about its behavior and properties can be formulated:

1. The PM-HLL adapts and locks to a periodic component emerging in the input signal in the PM-HLL catch range within a few repetitions of the signal.
2. Discontinuities in the signal periodicity are indicated by a drop in the HNR estimate. Locking is reached again after a few repetitions of the signal as long as the signal fundamental frequency stays in the PM-HLL catch range.
3. PM-HLL adaptation is ongoing once the PM-HLL is locked to a periodic component with a delay below the duration of one period.

4. The PM-HLL is noise-robust. Quick adaptation is even reached at low Signal-to-Noise Ratios (SNR) and with wide-band noise signals. Noise may include harmonic components at other fundamental frequencies.
5. The PM-HLL also adapts to quasi-periodic signals, including complex tones with shifted harmonics and signals with stochastic periodicity.

To test these hypotheses, simulations were performed using harmonic signals according to Equation (1) with additive $N(0,1)$–distributed Gaussian noise at different SNRs. Signals with different frequency content (fundamental missing or present) and with fixed, non-continuous and sweeping fundamental frequency tracks were used, which lie in the catch range of the PM-HLL instance. Furthermore, complex tones with shifted harmonics and a three-tone musical chord signal were tested. In the latter case, three PM-HLL instances were used in parallel to track each tone of the chord separately. Finally, Iterated Rippled Noise (IRN, [20]) was tested as an example of stochastic periodicity. IRN was generated by adding a 4-s sample of $N(0,1)$–distributed Gaussian noise and its delayed version in an iterative process (ADD-SAME procedure in [20]). Because the delay was implemented with a circular signal shift, a segment around the center of the resulting IRN was selected for processing to exclude edge effects. Short signals with a duration between 100 ms and 400 ms were used to check the high adaptation speed expected for the PM-HLL. The following simulations were performed, the details of which are listed in Table I:

I. The signal was a harmonic complex tone with a duration of 400 ms and a fixed $f_0$, which abruptly changed from 98.5 Hz to 101 Hz after 200 ms. The complex tone had five harmonics at different amplitudes and included the fundamental. Three different

SNRs were employed, 21.5 dB, 7.5 dB and 1.5 dB. Simulation I mainly tests hypotheses 1 and 2, but also 4.

II. Simulation II is the same as simulation I at a SNR of 7.5 dB, but all harmonics were shifted by +6 Hz or -6 Hz. A bias in the $f_0$-estimate is expected, but HNR should still remain rather high, and the variance in the $f_0$-estimate should not significantly increase compared to the harmonic case. Simulation II mainly tests hypothesis 5, but also 1, 2 and 4.

III. The signal in simulation III only contained the 6$^{th}$ and 7$^{th}$ harmonic, the SNR was 4.1 dB. Again, $f_0$ switched from 98.5 Hz to 101 Hz after 200 ms, with a total signal duration of 400 ms. Simulation III tests mainly hypothesis 1 for the special case of a missing fundamental.

IV. Simulation IV tests a complex tone with a linear sweep of its fundamental frequency from 96.0 Hz to 103.0 Hz within 100 ms. Again, harmonic numbers 1, 3, 4, 6 and 7 were included. Three SNRs, 21.3, 7.6 and 1.4 dB were tested. Simulation IV mainly tests hypotheses 3 and 4.

V. Simulation V tested an IRN with 5, 3 and 1 iterations to see whether the PM-HLL also adapts to stochastic periodicity (hypothesis 5). The fundamental frequency was 98 Hz.

VI. Simulation VI tested a chord signal with three tones at equal level, each including harmonic numbers 3, 4, 6 and 7. Furthermore, noise was added such that the SNR per component (against the noise plus the other two components) was -2.2 dB, and the total SNR of all three tones together was 3.1 dB. Each tone was estimated by a PM-HLL instance with an initial oscillator frequency at a semitone above the tone fundamental frequency. Simulation VI mainly tested hypothesis 4.

Table 1: List of simulations performed. 1st col.: Simulation number; 2nd col.: Brief signal description; 3rd col.: Total signal duration; 4th col.: Fundamental frequency; 5th col.: Composition of each complex tone (number of harmonic and its amplitude, n/a for IRN); 6th col.: Amount of mistuning of all components (simulation II only); 7th col.: Number if iterations of the IRN (simulation V only); 8th col.: Gain of additive $N(0,1)$-distributed Gaussian noise; 9th col.: Signal-to-Noise Ratio computed across the whole signal duration (n/a for IRN); 10th col.: Initial PM-HLL oscillator frequency: 11th col.: Tracking error in terms of mean and standard deviation across the whole signal duration, excluding the time period of initial convergence (cf. Results section).

|  | Signal | T / ms | $f_0$ / Hz | # of harmonic (amplitude) | $f_d$ / Hz | # | Noise gain | SNR / dB | Initial osc. freq. $f_{c,0}$ / Hz | Tracking error mean(std) / Hz |
|---|---|---|---|---|---|---|---|---|---|---|
| I | Complex tone with fixed fundamental frequency | 400 | 98.5 (1st half) 101.0 (2nd half) | #1(0.5),#3(0.9),#4(0.7), #6(0.9),#7(0.7) | 0 | - | 0.1, 0.5, 1.0 | 21.5, 7.5, 1.5 | 99.5 | 0.01 (0.37), 0.02 (0.59), 0.12 (0.81) |
| II | " | " | " | " | +6, -6 | - | 0.5 | 7.5 | " | +1.21 (0.59), -1.39 (0.71) |
| III | Complex tone with fixed fundamental frequency and missing fundamental | " | " | #6(0.9),#7(0.7) | 0 | - | 0.5 | 4.1 | " | 0.03 (0.56) |
| IV | Complex tone with linear sweep of its fundamental | 100 | 96.0 (start) 103.0 (end) | #1(0.5),#3(0.9),#4(0.7), #6(0.9),#7(0.7) | 0 | - | 0.1, 0.5, 1.0 | 21.3, 7.6, 1.4 | " | 0.45 (0.26), 0.44 (0.50), 0.90 (0.50) |
| V | Iterated Rippled Noise (IRN) | 200 | 98.0 | - | - | 5, 3, 1 | - | - | " | 0.00 (0.51), 0.15 (0.55), 0.60 (0.58) |
| VI | Major chord with three tones (in semitones: 1st, 4th and 7th) at equal level | 200 | 170.0 (tone 1), 214.2 (tone 2), 254.7 (tone 3) | For each tone: #3(0.9),#4(0.7), #6(0.9),#7(0.7) | 0 | - | 1.5 | -2.2 (one tone) +3.1 (three tones) | 183.6 (tone 1), 231.3 (tone 2), 275.1 (tone 3) | -0.7 (1.6), -1.0 (3.6), -0.5 (2.7) |

*Data analysis*

The estimation error was quantified by the mean and standard deviation across time of the difference between the PM-HLL oscillator frequency and the fundamental frequency of the signal $f_c(t) - f_0(t)$. The initial convergence part of the PM-HLL, which amounts to about 10% of the signal duration, was excluded from the statistics. Note that signals were rather short, the phases of the harmonics were selected at random, and the noise was random as well. The error values thus may vary when repeating the simulation. Yet, results show a clear pattern, so that further simulations were not required to test the hypotheses.

## 3. Results

Figure 3 shows the data of simulation I. SNR decreases from left to right panel, and each panel shows the HM-PLL oscillator frequency $f_c(t)$ referenced to 96 Hz (red curve, right scale), the HNR (blue curve, left scale) and the strobe points (black bars). The signal's fundamental frequency $f_0$ was +2.5 Hz (first half) and +5 Hz (second half) on this scale. Data show an initial convergence within about three periods, also after the change in the fundamental frequency, and an ongoing locking. The HNR increases quickly during adaptation, and drops within one period after the change in fundamental frequency, indicating an unlocked state. The HNR is higher than the nominal SNR due to the comb-filtering, and decreases with SNR. The statistics (last col. of Table 1, 1$^{st}$ entry) shows that mean and std. dev. of the tracking error increases with decreasing SNR, but remains in all cases well below 1 dB.

*** Place Fig. 3 here ***

The two left panels of Figure 4 show the same as simulation I for an SNR of 7.5 dB (center panel of Figure 3), but with a mistuning of all harmonics by +6 Hz (left panel) and -6 Hz (center panel). Data show the same pattern of loop convergence and HNR as in the harmonic case, with a similar tracking-error std. dev. of below 1 dB (last col. of Table 1, 2$^{nd}$ entry). This indicates that quasi-periodicity due to mistuning does not affect PM-HLL functionality, as expected. The fundamental frequency estimate during ongoing locking is biased by +1.21 Hz and -1.39 Hz, respectively, indicating that the mistuning systematically biases estimation.

*** Place Fig. 4 here ***

The right panel of Figure 4 shows the data for a complex tone with missing fundamental (simulation III). A pattern similar to the corresponding case of a complex tone with the fundamental present (center panel of Figure 3) is found. Note that SNR is by 3.4 dB lower than for the corresponding case, because the three lowest harmonics (1, 3 and 4) were discarded from the signal. Accordingly, the HNR drops by about the same amount. The statistics (last col. of Table 1, 3$^{rd}$ entry) shows that mean and std. dev. of the tracking error again remain well below 1 dB.

*** Place Fig. 5 here ***

Figure 5 shows the data of the sweep signal (simulation IV). SNR decreases from left to right panel, data presentation is as in Figure 3. Data clearly show quick convergence and an ongoing locking with a delay that is shorter than one period, even at the lowest SNR of 1.4 dB. The statistics (last col. of Table 1, 4$^{th}$ entry) again shows that mean and std. dev. of the tracking error increases with decreasing SNR, but remains in all cases below 1 dB.

*** Place Fig. 6 here ***

Figure 6 shows the data of the IRN (simulation V). The number of iterations was 5, 3 and 1 (left to right panels), data presentation is as in Figure 3. The estimate converges quickly to 98 Hz, which corresponds to the IRN delay. Mean and std. dev of the tracking error increases with less iterations, but stays in all cases well below 1 dB (last col. of Table 1, 5$^{th}$ entry). The HNR decreases with less iterations, and shows a higher variance as for the complex tones. It appears that the loop sometimes loses track and catches up again quickly, as indicated by the dips in the HNR, which coincide with a larger deviation in the fundamental frequency estimate.

*** Place Fig. 7 here ***

Figure 7 shows the data of the 3-tone major chord signal (simulation VI). The upper panel shows the fundamental frequency estimate of the three PM-HLL instances as a function of time. For each instance, the scale was referenced to the fundamental frequency of the tone it converges to. Although the SNR is at -2.2 dB for each of the tones, the PM-HLL instances converge within about 20 ms, which corresponds to about four periods of the lowest-frequency tone. The mean tracking error (last col. of Table 1, 6[th] entry) is at or below 1Hz. The std. dev. of the tracking error is higher than in the other simulations, but it is still rather low, given the high level of noise. The panel also shows the strobe points of each PM-HLL instance, with shorter bar lengths for higher tones. The middle panel of Figure 7 shows the mean HNR per component, which is around 5 dB, i.e., about 7 dB higher than the signal SNR. The lower panel shows the non-normalized intensity spectrum of the first 30 ms of the signal (coarse curve) and of the complete signal (fine curve). The spectrum of the first 30 ms does not show any clear tonal components, i.e., it appears impossible to estimate the chord from this FFT data. The spectrum of the complete signal would allow a chord estimate, but some logic would be needed to derive the correct tones and associate the tonal components. Note that the PM-HLL converged within 20 ms.

## 4. Discussion

The simulations using a variety of test signals showed that the PM-HLL adapts to periodic signal components emerging in a wideband time signal within 3-4 periods of the onset of the component and tracks changes in the component's fundamental period, or fundamental frequency, with a delay that is well below the duration of one period. To the author's knowledge, similarly short adaptation and tracking time constants have not been achieved with established methods, at least not for SNR values at or even below 0 dB. This achievement may be due to the sensitivity of the novel phase-based control signal, which is directly derived from the model of a harmonic signal with varying fundamental frequency.

*\*\*\* Place Fig. 8 here \*\*\**

To illustrate a possible application of the PM-HLL to speech signals, a signal with time-varying values of fundamental frequency, first and second formant was generated using the vocoder method introduced in [28]. The signal sounds like a continuous succession of different vowels with varying pitch. This type of signal was used in [28] to investigate human attentive tracking of voices and was therefore considered suitable for the purpose of demonstrating pitch tracking using the PM-HLL[6]. Figure 8 shows the spectrogram of the signal using a 512-tap windowed FFT with a window overlap of 90%. The signal's initial fundamental frequency is at about 245 Hz. The left panel shows the spectrogram of the signal at an SNR of 0 dB and the fundamental frequency estimate of a PM-HLL instance that was started at an oscillator frequency $f_c$ of 310 Hz (black curve). The right panel shows the spectrogram of the clean voice and again the PM-HLL estimate, together with its integer multiples 2-9. The PM-HLL instance catches the fundamental frequency trajectory at about 100 ms into the signal and keeps track throughout the duration of the signal. The precision of the estimate can be best assessed at the higher harmonics. The deviation at the 8$^{th}$ harmonic is at or below about 3%. Even the rapid changes at about 0.6 s and 1.0 s are tracked. Note that the spectrogram of the noisy signal, from which the PM-HLL estimate was derived, does not show the fundamental, and most of the time only the second harmonic is visible.

---

[6] The signal is included in the supplementary material [18]. After listening to the signal, a human can reproduce it. It is thus very similar to a human voice, but has the advantage that the ground-truth tracks of the fundamental frequency and of the two formants are known. Code for stimulus generation is publicly available at [29].

A requirement for the algorithm to adapt and converge is that the emerging periodic component lies in the catch range of the PM-HLL instance. This is the case for the example in Figure 8, although the initial values of the signal's fundamental frequency and of the PM-HLL instance's oscillator frequency were quite apart. To scan for a whole range of potentially observable fundamental frequencies in parallel, a set of PM-HLL instances with different initial oscillator frequencies is required, all sharing the same delay line. The data from such a set, i.e., the tracks of oscillator frequency and HNR for each instance, can be sent upstream for further analysis and interpretation, e.g., by a Computational Scene Analysis (CASA) algorithm or a multi-pitch tracker. The tracks with salient periodicity (i.e., HNR is high) form segregated periodic glimpses of the sound, of which streams can be formed, e.g., for tracking speech. If only one instance of the set indicates a salient periodicity, only one source is active at that point in time, as it is the case for the signal from Figure 8, and the subsequent analysis is rather easy. If multiple salient periodicities are detected at the same point in time, a multi-source scenario is detected. Consequently, the subsequent analysis stage must form multiple streams to segregate the different sources. In that sense, a set of PM-HLL instances provides a set of periodicity-related features that may replace, e.g., an autocorrelation analysis, if low-delay processing with high temporal resolution is required. The advantage is that multiple periods can be detected in parallel, which allows the subsequent processing to segregate different sources.

Depending on the application, about 7-12 PM-HLL-instances may be sufficient to monitor a full octave of fundamental frequencies, depending on the signal's bandwidth. To monitor three octaves, about 30 PM-HLL instances are needed. However, periodic signals are also periodic at integer multiples of the period (sub-harmonics), i.e., PM-HLL instances can be coupled across octaves. Although each PM-HLL instance is low in computational effort, the effort scales with the number of instances used. Note, however, that an established algorithm like YIN [27] performs a greedy search across all possible periods in each time frame. To scan the full range of three octaves from 50 Hz to 400 Hz at a sampling rate of 5 kHz, for example, about 90 different periods must be tested (delays between 10 and 100 samples). For each period and each frame, a measure related to the power of the output of the period-suppressive comb filter (2$^{nd}$ line of Equation 5) is computed, and the period is selected that is associated with the minimum value of the power measure. This means that YIN is comparable to a set of PM-HLL instances in terms of computational effort, but lacks the sensitivity and time-resolution of the phase-based adaptation and is by definition not capable of finding multiple periods in parallel.

The application of the PM-HLL is more straightforward if the targeted value or range of fundamental frequencies is known. Assuming that prior knowledge is available at a certain instance of time on which fundamental frequencies may emerge, e.g., from a predictive CASA algorithm, new PM-HLL instances may be initialized at these frequencies. The number of required PM-HLL instances may be reduced then. Combinations with CASA algorithms will be investigated in future studies.

Although not tested in this study, the results suggest that the PM-HLL can be applied to frequency sub-band signals from a filterbank or from short-time Fourier transform (STFT).

For STFT-based filterbanks as well as other modulating filterbanks, a phase reconstruction may be required by modulating all frequency bands into the baseband (eqn. 4 in [21]).

To properly represent the signal waveform while being able to follow changes in the signal period, all time constants must be selected to reflect the rate of change of the fundamental frequency that occurs in the signal. Speech, e.g., includes rapid changes in the fundamental frequency, i.e., time constants must be rather short, thereby limiting the effective number of period-averages to only a few periods. For music, the time constants may be longer. In this study, time constants were chosen to be a fixed (small) multiple or fraction of the current PM-HLL oscillator period. Any options for its automatic adaptation to the input's characteristics may be investigated in a subsequent study.

Note that all simulations of this study computed the PM-HLL control signal without further processing. By design of the adaptive loop control, the control signal must oscillate around the true fundamental frequency. Tracking algorithms such as Kalman filters (e.g. [22]) or particle filters (e.g. [23]) may be used to stabilize the estimate without adding significant delay.

The fact that all time constants are related to the current PM-HLL oscillator frequency makes the algorithm fully scalable to other frequency and fundamental frequency ranges, as long as the sampling frequency is adapted properly. This is an important feature that makes the algorithm potentially interesting for a range of applications, in particular if low-delay processing is required.

The PM-HLL may be applied to extending *auditory models* of fine-grain temporal processing. It can be considered an extension to the strobed-integration model [17] in that it adds a reliable strobe-point generation mechanism. All results obtained with the auditory strobed-integration model thus also apply to the PM-HLL. The PM-HLL can also be considered an extension of the cancellation model [16] in that it simplifies the search for cancellation delays. The results obtained with the cancellation model thus also apply to the PM-HLL[7].

Another application area may be *audio signal processing*, in particular for devices that require low-delay processing. Hearing aids, e.g., require the total delay to be well below 10 ms, which makes the application of pitch-based processing difficult but yet interesting (e.g. [21]). The PM-HLL may allow a new perspective on pitch-based processing for hearing aids. In particular, it may be used for cancelling unwanted periodic components and/or enhancing desired target-related periodic speech components. Another option is to improve spectral estimates by using the PM-HLL to derive period-synchronous ("pitch-synchronous") signal features such as the spectral profile or the Harmonic-to-Noise Ratio (HNR). Although not tested in this study, the results suggest that a period-synchronous spectral analysis may be performed on the period-synchronous features to estimate the spectro-temporal profile of the periodic component the PM-HLL is locked to.

The PM-HLL may also augment algorithms for *line-artifact suppression* for electrophysiological experiments (e.g. [25]) by providing a precise estimate of the power of the fundamental and harmonics of the line artifact for optimal suppression. Note that the period-suppressive comb filter (eqn. 5) suppresses all harmonics of the component the PM-HLL is locked to.

---

[7] An important characteristic of auditory peripheral processing is the demodulation of the beating of harmonics falling into the same auditory frequency bands by the inner haircell function, a basic model of which is half-wave rectification and lowpass filtering. Periodicity can therefore be detected from resolved (single) harmonics at lower frequencies, and at the fundamental and potentially the first harmonic at higher frequencies [24]. Such a specific demodulation may also be interesting for audio signal processing applications.

Although rather speculative, the PM-HLL may also be used for *high-frequency applications* such as radio receivers. For this, it may be required to use tunable analog delay lines (e.g. [26]) for implementing the comb filters (eqn. 5).

Another speculative application is *power electronics* and *grid synchronization* [6], because here, specific PLLs are used, which may not be optimal for estimating line signals with unknown phase and amplitude values of its harmonics.

## 5. Conclusions

The simulation data confirm the theoretical expectation, that the Period-Modulated Harmonic-Locked-Loop (PM-HLL) enables rapid and noise-robust $f_0$ (fundamental frequency) estimation from wide-band periodic signals in the time-domain. In particular, the following conclusions can be drawn:

1. Rapid convergence: A PM-HLL instance converges within 3-4 periods from the signal onset, even at low SNRs around 0dB.

2. Quasi-periodicity: The PM-HLL also tracks quasi-periodic signals, such as complex tones with shifted harmonics and signals with stochastic periodicity.

3. Sub-period sensitivity: Once locked to a periodic component, a PM-HLL instance follows a continuously changing $f_0$ with an estimation delay of one period or below.

4. Multi-periodicity: A PM-HLL instance only adapts and locks to emerging periodic signal components, if its fundamental frequency lies within an interval around the current oscillator frequency of the PM-HLL instance. This "catch range" depends on the signal bandwidth and number of harmonics. For typical applications, around ten parallel PM-HLL instances are sufficient to monitor a full octave of fundamental frequencies.

5. The PM-HLL is the first algorithm to combine additive and subtractive (cancellation) models of periodicity processing to derive a phase-based loop control signal. Its adaptation speed and sensitivity is due to the direct application of a time-domain harmonic signal model to the estimation problem.

6. Parallel and/or sequential estimation of temporally-overlapping periodic components using several PM-HLL instances may facilitate separation of concurrent sources in scene analysis algorithms.

## Acknowledgements

Funded by the Deutsche Forschungsgemeinschaft (DFG, German Research Foundation) – Project-ID 352015383 – SFB 1330 – project B2.

## Data Availability Statement

All simulation data and figures presented here can be computed with Matlab scripts that are available at [18].

# Figure Captions

Figure 1: Principle of strobe-point generation: The fundamental frequency estimate of a periodic signal is assumed to increase linearly from 50 Hz to 150 Hz in the time interval 0 ms to 100 ms (blue curve, left scale). The corresponding phase function according to (2) is normalized by $2\pi$ and taken modulo 1 (red curve, right scale). Whenever the phase wraps, a strobe point is set (indicated by black bars with circles on top). The time span between subsequent strobe points indicates the stabilized period.

Figure 2: Filter transfer functions in the range between $f_c/2$ and $3f_c/2$, centered at the current PM-HLL oscillator frequency (i.e., fundamental frequency estimate) $f_c$. The amplitude is shown in blue (left scale), and the normalized phase is shown in red curves (right scale). The period-constructive comb filter (solid lines) enhances a component at $f_c$, whereas the period-suppressive comb filter (dashed lines) cancels it. If the true fundamental frequency $f_0$ of a signal is lower (indicated as $f_{0,1}$) or higher (indicated as $f_{0,2}$) than $f_c$, the phase difference between the two comb filters is $+90°$ or $-90°$, respectively. This phase difference is used by the PM-HLL algorithm as a control signal to adapt the fundamental frequency estimate $f_c$ towards $f_0$. Note that, due to the periodicity of the transfer functions involved, the same behavior applies to all harmonics of the signal, i.e., the phase flip between $+90°$ and $-90°$ occurs at the same value of $f_c$ for the fundamental and all its harmonics. Thus, the fundamental and all its harmonics contribute to the control signal in the same way.

Figure 3: Simulation I: Complex tone with a fixed fundamental frequency of 98.5 Hz (1st half) and 101.0 Hz (2nd half), SNR is 21.5 dB, 7.5 dB and 1.5 dB (left, center and right panel). Each panel shows the PM-HLL oscillator frequency (i.e., fundamental frequency estimate) $f_c$ referenced to 96 Hz (red curve, right scale), the HNR (blue curve, left scale) and the strobe points (black bars), as a function of time. The black line shows the signal's fundamental frequency trajectory. Data show quick convergence and small estimation variance even at the lowest SNR. The HNR depends on SNR and quickly goes low during loop adaptation, thus indicating the unlocked state.

Figure 4: Complex tone with all harmonics mistuned by +6 Hz (left panel) and -6 Hz (center panel) at a SNR of 7.5 dB (Simulation II; same as center panel of Figure 3, but with mistuning). Data presentation is the same as in Figure 3. Mistuning does not affect PM-HLL functionality, but the fundamental frequency estimate is biased on average by +1.21 Hz and -1.39 Hz, respectively. Right panel: Data for a complex tone with missing fundamental (simulation III) shows a similar pattern as the corresponding case of a complex tone with the fundamental present (center panel of Figure 3). The HNR drops by about 3.4 dB compared to the corresponding case, because the three lowest of in total five harmonics were discarded from the signal.

Figure 5: Complex tone with a linear sweep of the fundamental frequency from 96 Hz to 103 Hz (simulation IV). SNR decreases from left to right panel, data presentation is as in Figure 3. Data clearly show quick convergence and an ongoing locking with a delay that is shorter than one period, even at the lowest SNR of 1.4 dB (right panel). Note that the total duration of the signal is only 100 ms.

Figure 6: IRN with 5, 3 and 1 iterations (left to right panels, simulation V). Data presentation is as in Figure 3. The estimate converges quickly to the target value of 98 Hz (+2Hz on the relative frequency scale). The HNR decreases with less iterations and shows some variance, but the fundamental frequency estimate is still quite accurate.

Figure 7: 3-tone major chord signal (simulation VI). The upper panel shows the oscillator frequency (i.e., fundamental frequency estimate) $f_c$ of three PM-HLL instances as a function of time. Each instance targeted one of the tones by selecting the instance's initial $f_c$ to be a semitone above the target's $f_0$, i.e., $f_0$ lies at the upper edge of the instance's catch range. In the plot, each instance's $f_c$ time course was referenced to its target $f_0$, i.e., the target $f_0$ is plotted at 0 Hz. The PM-HLL instances converge within about 20 ms at an SNR of -2.2 dB for each tone. Strobe points are plotted separately for the different instances, with shorter bar lengths for higher tones. The middle panel shows the mean HNR per component, and the lower panel shows the non-normalized intensity spectrum of the first 30 ms of the signal (coarse curve) and of the complete signal (fine curve). The tones may not be derived from the 30-ms FFT data, because tonal components are not resolved. Note that the PM-HLL converged in 20 ms.

Figure 8: Spectrogram of a vocoded voiced speech signal with time-varying trajectories of fundamental frequency, first and second formant. The left panel shows the spectrogram of the signal at an SNR of 0 dB as well as the fundamental frequency estimate derived from the noisy signal with a PM-HLL instance that was started at an oscillator frequency $f_c$ of 310 Hz (black curve). The right panel shows the spectrogram of the clean voice and again the PM-HLL estimate, together with its integer multiples 2-9. The PM-HLL instance catches the fundamental frequency trajectory at about 100 ms into the signal and keeps track throughout the duration of the signal. Note that the spectrogram of the noisy signal does not show the fundamental, and most of the time only the second harmonic is visible.

Figures:

Figure 1:

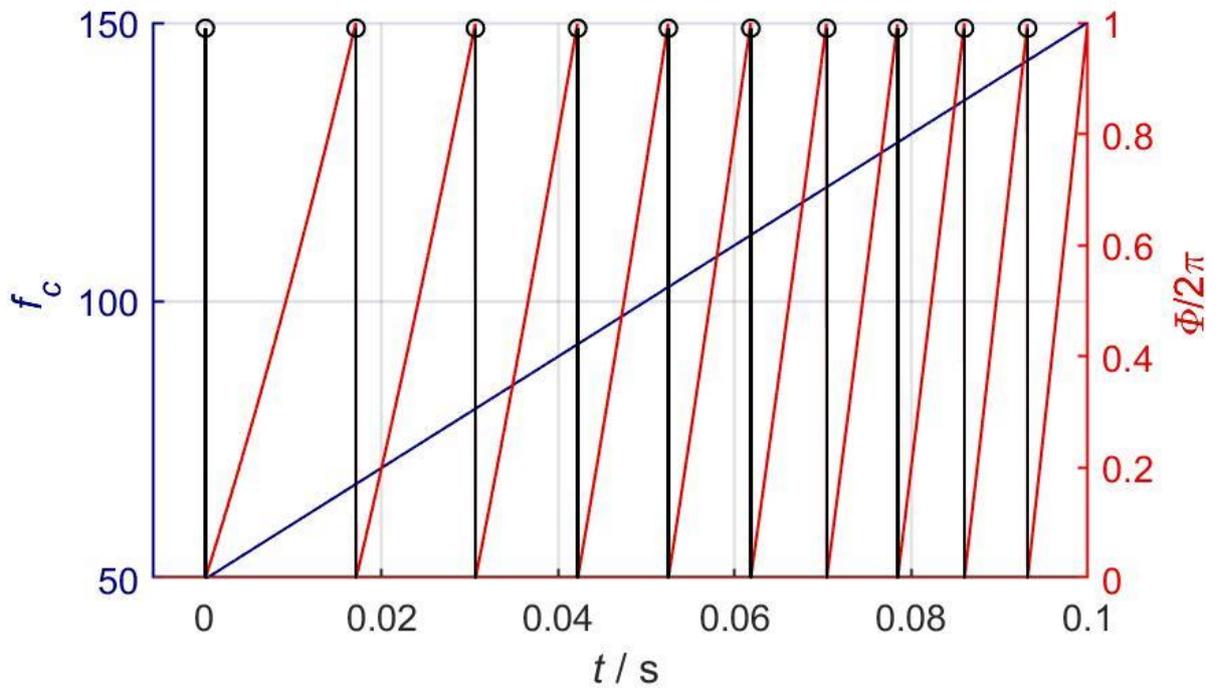

Figure 2:

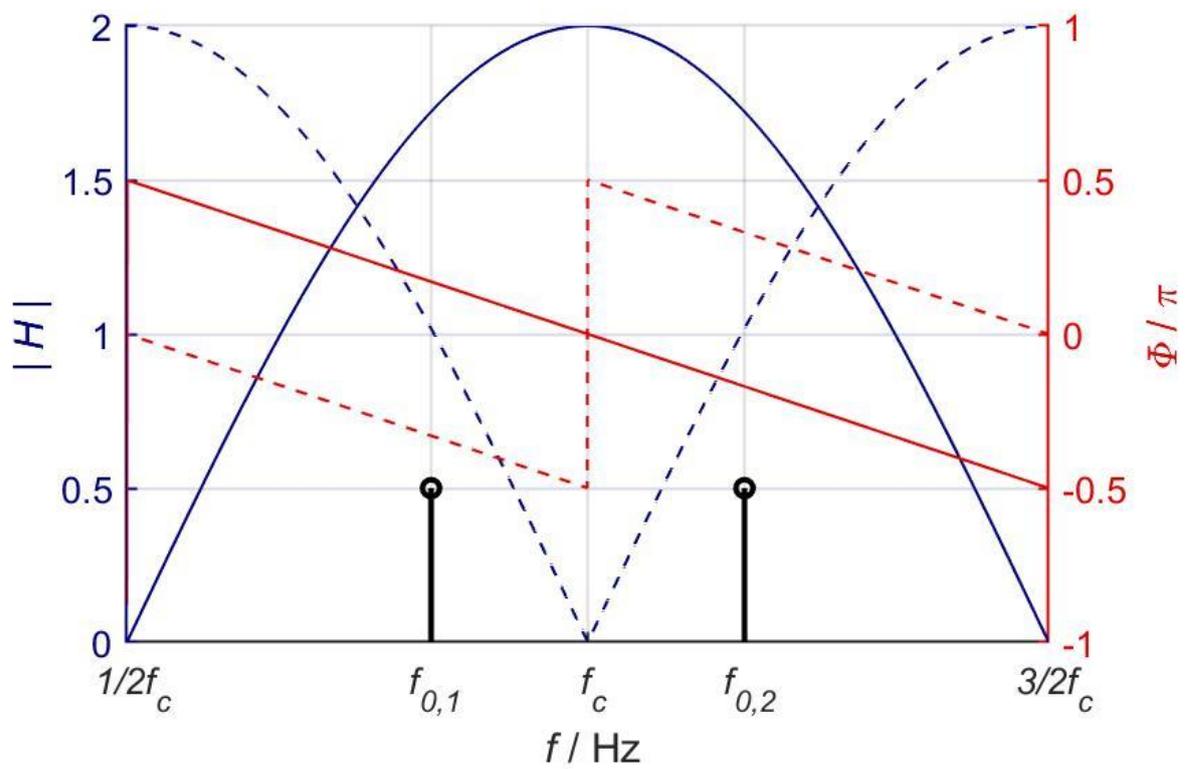

Figure 3:

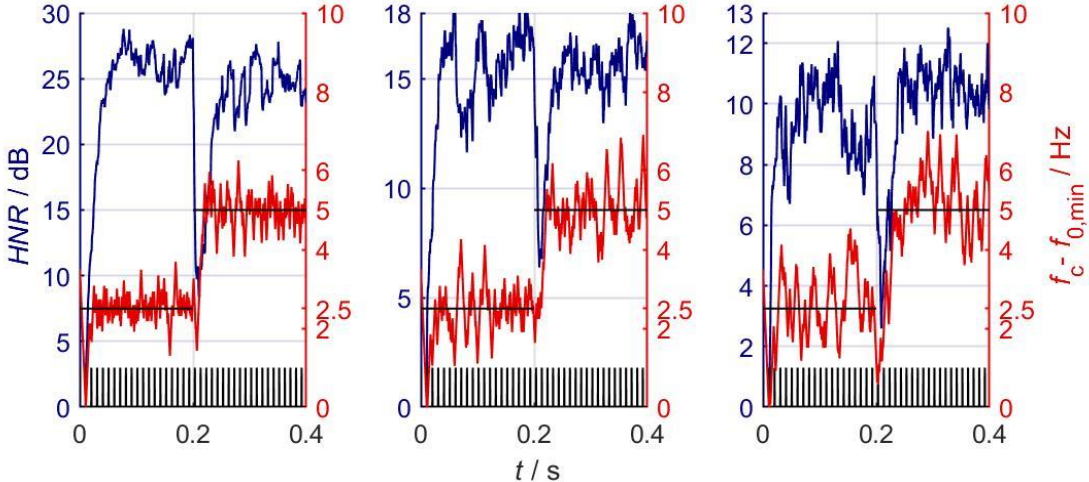

Figure 4:

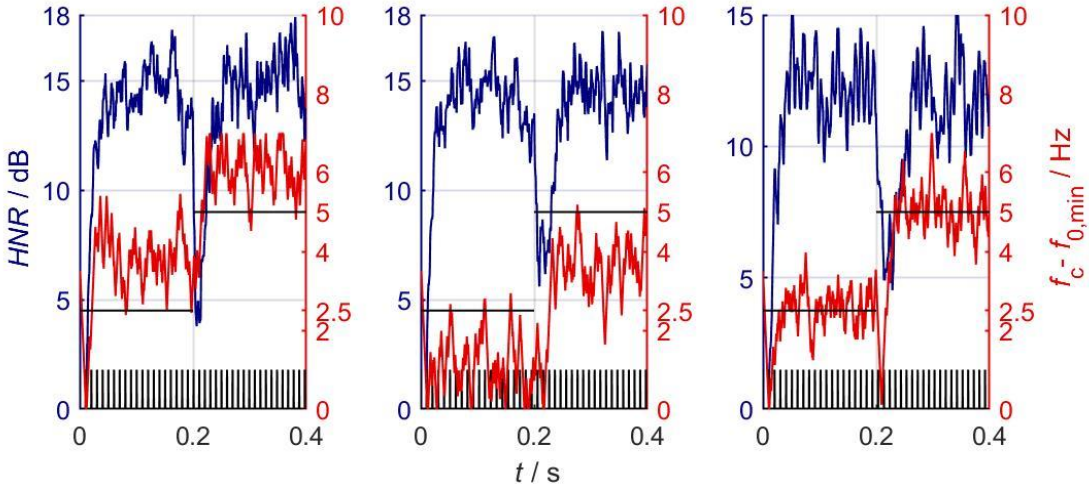

Figure 5:

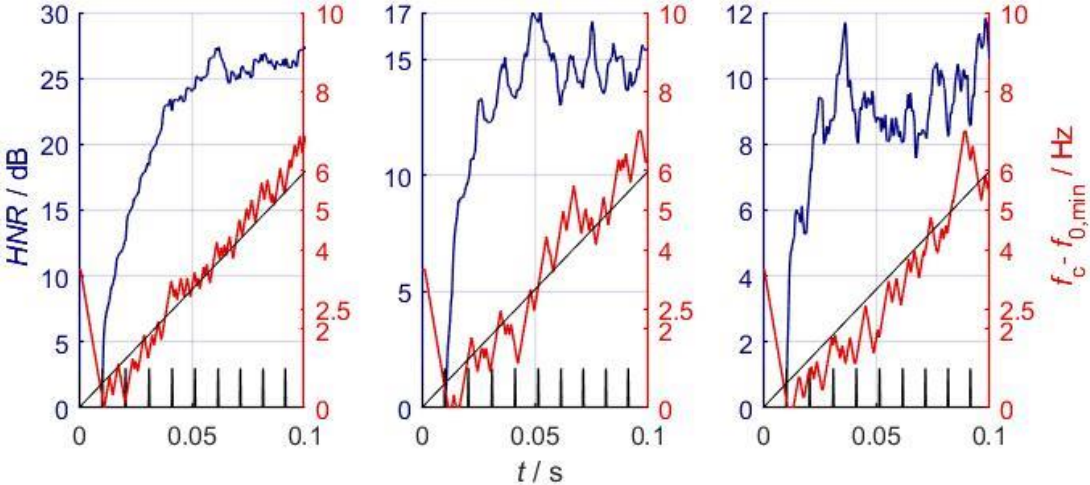

Figure 6:

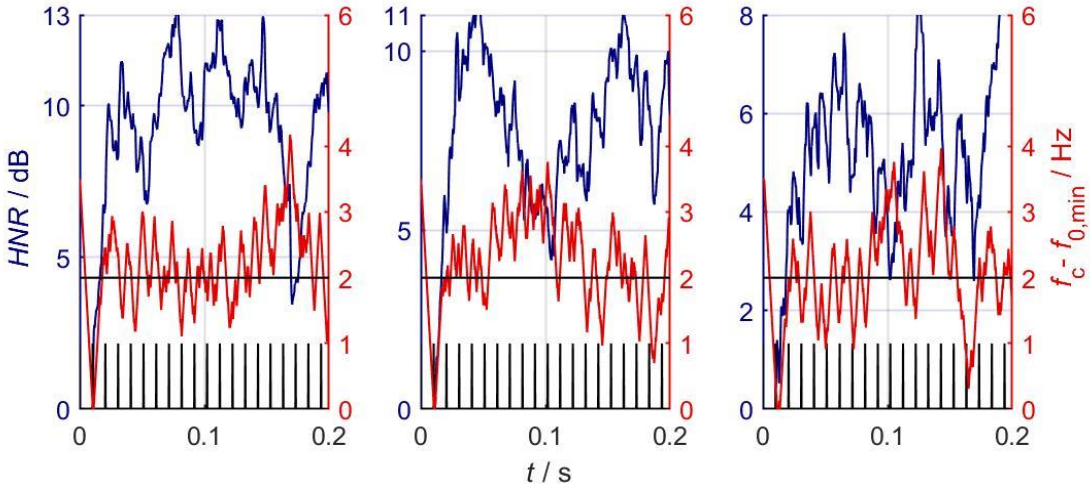

Figure 7:

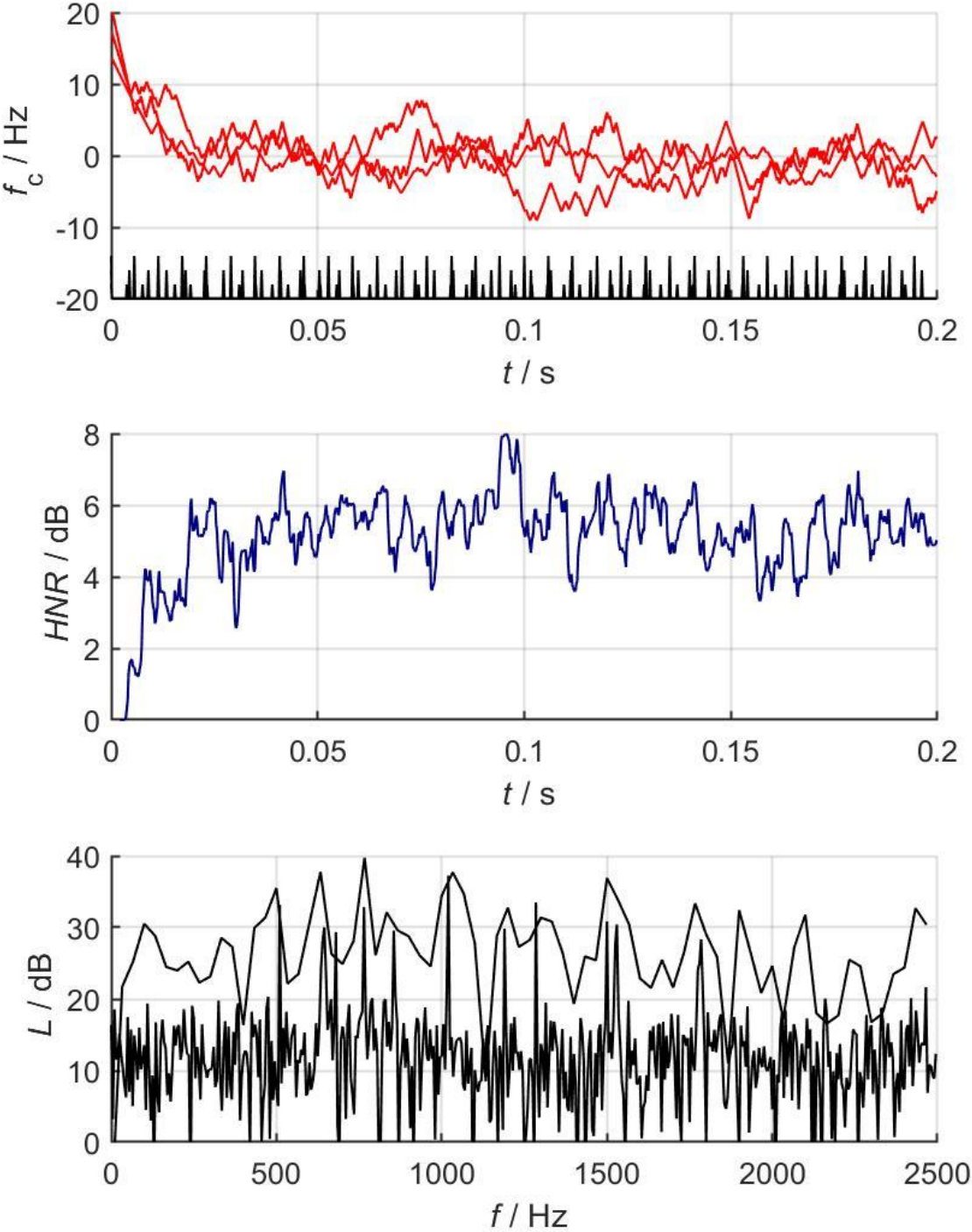

Figure 8:

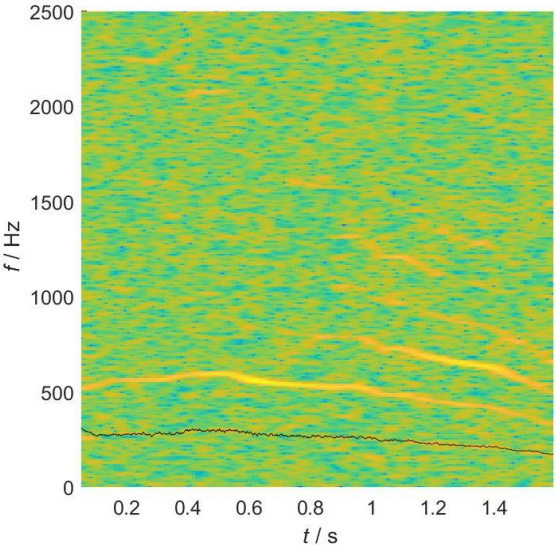 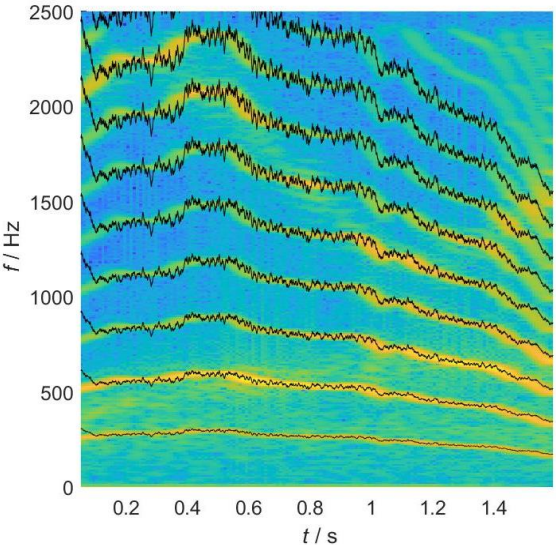